\documentclass[a4paper,12pt]{article}
\usepackage{amsmath}\usepackage{amssymb}
\usepackage{latexsym}\usepackage{cite}

\usepackage{amsmath}
\usepackage[dvips]{graphicx}
\usepackage{pstricks}
\usepackage{color}
\usepackage{bm}

\usepackage{graphicx}
\usepackage{caption}
\usepackage{varioref}
\usepackage{placeins}
\usepackage{float}
\usepackage{amssymb}
\usepackage{amsmath}
\usepackage{graphicx}
\usepackage{verbatim}
\usepackage{color}
\usepackage{subfigure}
\usepackage{hyperref}
\usepackage[final]{pdfpages}
\usepackage{afterpage}
\usepackage{makecell}

\definecolor{Red}{rgb}{1.00, 0.00, 0.00}
\definecolor{Green}{rgb}{0.00, 1.00, 0.00}
\definecolor{Blue}{rgb}{0.00, 0.00, 1.00}
\definecolor{Cyan}{rgb}{0.00, 1.00, 1.00}
\definecolor{Mymagenta}{rgb}{0.3, 0.00, 1.00}%
\definecolor{Magenta}{rgb}{1.00, 0.00, 1.00}
\definecolor{DeepSkyBlue}{rgb}{0.00, 0.75, 1.00}
\definecolor{DarkGreen}{rgb}{0.00, 0.39, 0.00}
\definecolor{SpringGreen}{rgb}{0.00, 1.00, 0.50}
\definecolor{Mygreen}{rgb}{0.00, 0.72, 0.00}
\definecolor{DarkOrange}{rgb}{1.00, 0.55, 0.00}
\definecolor{OrangeRed}{rgb}{1.00, 0.27, 0.00}
\definecolor{DeepPink}{rgb}{1.00, 0.08, 0.57}
\definecolor{DarkViolet}{rgb}{0.58, 0.00, 0.82}
\definecolor{SaddleBrown}{rgb}{0.57, 0.27, 0.07}
\definecolor{Black}{rgb}{1.00, 1.00, 1.00}
\definecolor{Ablue}{rgb}{0.10, 0.1, 1.00}

\def\beq{\begin{equation}}
\def\eeq{\end{equation}}
\def\la{\label}
\def\beqs{\begin{equation*}}
\def\eeqs{\end{equation*}}

\def\beq{\begin{equation}}
\def\eeq{\end{equation}}
\def\beqr{\begin{eqnarray}}
\def\eeqr{\end{eqnarray}}

\def\lam{\lambda}

\def\la{\label}

\def\vs{\vspace}

\allowdisplaybreaks
\begin{document}
\begin{flushright}
March 22, 2016 \\
\end{flushright}

\vs{1.5cm}

\begin{center}
{\Large\bf

 Neutrino Mass Matrices from Two Zero \\

 \vs{0.3cm}

$3\times 2$ Yukawa Textures and Minimal $\rm d=5$ Entries
 }

\end{center}

\vspace{0.5cm}
\begin{center}
{\large
~Avtandil Achelashvili\footnote{E-mail: avtandil.achelashvili.1@iliauni.edu.ge},~ and
~Zurab Tavartkiladze\footnote{E-mail: zurab.tavartkiladze@gmail.com}
}
\vspace{0.5cm}

{\em Center for Elementary Particle Physics, ITP, Ilia State University, 0162 Tbilisi, Georgia}
\end{center}

\vspace{0.6cm}

\begin{abstract}


Aiming to relate leptonic CP violating phase $\delta$ to the cosmological CP asymmetry,
we study the extension of MSSM by two quasi-degenerate (strictly degenerate at tree level) right- handed neutrinos and consider all possible two texture zero $3\times2$ Yukawa matrices plus one $\Delta L=2$ dimension five ($\rm d=5$) operator contributing to the light neutrino mass matrix.
We classify all experimentally viable mass matrices, leading to several predictions, and analytically derive predictive relations. We also relate the CP violating $\delta$ phase to the CP phase of the thermal leptogenesis.

\end{abstract}

\hspace{0.4cm}{\it Keywords:}~Neutrino masses and mixings; leptogenesis.

\hspace{0.4cm}{\rm PACS numbers:}~11.30.Fs, 12.60.Jv, 14.60.St, 98.80.-k

\newpage

\section{Introduction}
\label{intro}

Although it's great success, the standard model (SM) needs some extension. In order to accommodate
atmospheric and solar \cite{Fogli:2012ua},\cite{Gonzalez-Garcia:2014bfa} neutrino data, the neutrino masses and mixings should be generated via some
reasonable extension.
See-saw mechanism \cite{{Minkowski:1977sc},{GellMann:1980vs},{Yanagida:1979as},{Glashow:1979nm},{Mohapatra:1979ia},{Schechter:1980gr},{Schechter:1981cv}} realized by the introduction of the heavy right-handed neutrinos (RHN), is simplest one for neutrino mass generation.
Additional an appealing feature of this extension is that it can also generate the needed amount of the baryon asymmetry
via leptogenesis \cite{Fukugita:1986hr} (for reviews see: \cite{{Giudice:2003jh},{Buchmuller:2004nz},{Davidson:2008bu}}).
Since the neutrino sector involves CP phases and parameters (e.g. Dirac Yukawa couplings and heavy Majorana neutrino masses) which are not measured so far, a priory it is impossible
to make predictions unless some reduction of model parameters are achieved.
For this purpose, the texture zero Yukawa and/or Majorana mass matrices have been investigated in the literature \cite{{Frampton:2002qc},{Frampton:2002yf},{Ibarra:2003up},{Branco:2005jr},{Fritzsch:2011qv},{Dev:2006qe},{Xing:2002ta},{Grimus:2011sf},{Dev:2014dla},{Zhou:2015qua},{Kitabayashi:2015jdj},{Meloni:2014yea},{Merle:2006du},{Lashin:2011dn},{Deepthi:2011sk},{Liao:2013rca},{Gautam:2015kya},{Nath:2015emg}}.
This approach, besides some predictions, opens up a possibility of relating
 phase $\delta $ (appearing in neutrino oscillations) to the CP asymmetry of the thermal leptogenesis \cite{{Frampton:2002qc},{Frampton:2002yf},{Ibarra:2003up}},\cite{{Babu:2007zm},{Babu:2008kp},{Harigaya:2012bw},{Ge:2010js}}.

Since for a solution to the gauge hierarchy problem the supersymmetry appears to be a well motivated
framework, we consider the MSSM augmented with two RHN states. The latter being quasi-degenerate in mass
have potential to realize a resonant leptogenesis scenario \cite{{Flanz:1996fb},{Pilaftsis:1997jf},{Pilaftsis:2003gt}} (and \cite{{Blanchet:2012bk},{Dev:2015wpa},{Dev:2014laa},{Dev:2014wsa}} for recent discussions on resonant leptogenesis) which would not suffer from the gravitino problem \cite{{Khlopov:1984pf},{Ellis:1984er},{Davidson:2002qv},{Kohri:2005wn}}.
Noting also that the low scale SUSY has the dark matter candidate,
the framework we are considering, is well motivated from the several viewpoints.

With the two  RHN's we investigate  texture zero $3\times 2$ Dirac type Yukawa couplings, which lead to the neutrino mass
matrices with zero entries. On top of this, we augmented the Lagrangian couplings  with a single $\Delta L=2$  lepton number violating $\rm d=5$ operator, which
allows to keep some predictions and, at the same time, makes some mass matrices experimentally acceptable.
  It turns out
that only three Yukawa textures (out of nine) possess cosmological CP phase which we relate to
neutrino CP  $\delta$ phase. All experimentally viable neutrino mass matrices lead to interesting predictions, which we investigate
in detail.

The paper is organized as follows. In section \ref{matrices}, we describe
our framework and list all possible two texture zero $3\times 2$ Yukawa matrices. In section \ref{derivation},
resorting to the $\rm d=5$ operator and $3\times 2$ Yukawa matrices we construct neutrino mass matrices.
Simple example of possible generation of $\rm d=5$ operators, we are exploiting, is also outlined.
In section \ref{analysis}, parametrization of the lepton mixing matrix is given and experimentally acceptable mass matrices
are recognized. We investigate these neutrino mass matrices and derive predictive relations, some of which are exact and very applicable to analysis.
In section \ref{cosmology}, cosmological CP phase is related to the $\delta $ phase responsible
for the CP violation in neutrino oscillations. In Sect. \ref{conclusion} we conclude.
\section{Two texture zero $3\times2$
Yukawa matrices: $2T_0Y_{32}$'s}
\la{matrices}
\numberwithin{equation}{section}
Let us consider the lepton sector of MSSM augmented with two
right-handed neutrinos $N_{1}$ and $N_{2}$. The relevant
Yukawa superpotential couplings are given by:
\begin{equation}
W_{lept}=W_{e}+W_{\nu},\quad W_{e}=l^{T}Y_{e}^{\rm diag}e^{c}h_{d},\quad
W_{\nu}=l^{T}Y_{\nu}Nh_{u}-\frac{1}{2}N^{T}M_{N}N \la{r21},
\end{equation}
where $h_{d}$ and $h_{u}$ are down and up type MSSM Higgs doublet
superfields respectively. $N$, $l$, $e^{c}$  denote:
\begin{equation}
N=\binom{N_{1}}{N_{2}}, \quad l^{T}=(l_{1}, l_{2}, l_{3}), \quad
e^{cT}=(e^{c}_{1}, e^{c}_{2}, e^{c}_{3}).
\end{equation}
In the next section, upon deriving the neutrino mass matrices, together with couplings of Eq. (\ref{r21}), the single $\rm d=5$ operator per the neutrino mass matrix will be applied. Because of this, in comparison with the approach considered in \cite{Babu:2008kp}, more two texture zero $Y_{\nu }$ Yukawa matrices will be compatible with the current experiments. We will work in a basis in which the charged lepton Yukawa matrix
is diagonal and real:
 \beq
 Y_{e}^{\rm diag}={\rm Diag}(\lambda_{e}, \lambda_{\mu}, \lambda_{\tau}).
\eeq
 As far as the RHN mass matrix $M_{N}$ is concerned, we will
assume that
it has the form:
 \beq
M_{N}= \left(\begin{array}{ccc}
 0&1\\
 1&0
\end{array}\right)M. \la{m01}
\eeq
 This form of $M_{N}$ is crucial for our studies, since (\ref{m01}) at a tree level leads to the mass degeneracy of the RHN's, it has
interesting implications for resonant leptogenesis
\cite{Babu:2007zm},\cite{Babu:2008kp} and also, as we will see
below, for building predictive neutrino scenarios. In a spirit of \cite{Babu:2008kp}, here we attempt
to classify specific texture zero scenarios with degenerate RHN's which lead
to predictions consistent with experiments.
 The matrix $Y_{\nu}$ contains two columns. Since due to the form
 of $M_{N}$ there is an exchange invariance $N_{1}\rightarrow N_{2}$,
 $N_{2}\rightarrow N_{1}$, it does not matter in which column of
 $Y_{\nu}$ we set elements to zero. Thus, starting with the Yukawa couplings, we consider the following nine different $3\times2$ Yukawa matrices with two zero entries:
\\
\beqs
T_{1}= \left(\begin{array}{ccc}
\times&0\\
\times&0\\
\times&\times
\end{array}\right),\quad
T_{2}= \left(\begin{array}{ccc}
\times&0\\
\times&\times\\
\times&0
\end{array}\right),\quad
T_{3}= \left(\begin{array}{ccc}
\times&\times\\
\times&0\\
\times&0
\end{array}\right),
\eeqs
\beqs
T_{4}= \left(\begin{array}{ccc}
0&0\\
\times&\times\\
\times&\times
\end{array}\right),\quad
T_{5}= \left(\begin{array}{ccc}
\times&0\\
0&\times\\
\times&\times
\end{array}\right),\quad
T_{6}= \left(\begin{array}{ccc}
\times&0\\
\times&\times\\
0&\times
\end{array}\right),
\eeqs
\beq
T_{7}= \left(\begin{array}{ccc}
\times&\times\\
0&0\\
\times&\times
\end{array}\right),\quad
T_{8}= \left(\begin{array}{ccc}
\times&\times\\
\times&0\\
0&\times
\end{array}\right),\quad
T_{9}= \left(\begin{array}{ccc}
\times&\times\\
\times&\times\\
0&0
\end{array}\right),\la{xxx}
\eeq
 where "$\times$"s stand for non-zero entries. Next, we factor
out phases from these textures, in such a way as to make maximal
number of entries be real. As it turns out only $T_4, T_7$ and
$T_9$ will have unfactorable phases. The latter should be relevant
to the lepton asymmetry.
\\
\\
TEXTURE $T_{1}$
\\
Starting with $T_{1}$ Yukawa matrix, we parameterize  it and write in a form of factored out phases:
\beq
T_{1}=\begin{pmatrix}
a_{1}e^{i\alpha_{1}} & 0\\
a_{2}e^{i\alpha_{2}} & 0\\
a_{3}e^{i\alpha_{3}} & b_{3}e^{i\beta_{3}}
\end{pmatrix}
=
\begin{pmatrix}
e^{ix} & 0&0\\
0 & e^{iy}&0\\
0 & 0&e^{iz}
\end{pmatrix}
\begin{pmatrix}
a_{1} & 0\\
a_{2} & 0\\
a_{3} &b_{3}
\end{pmatrix}
\begin{pmatrix}
e^{i\omega} & 0\\
0 & e^{i\rho}
\end{pmatrix},
\eeq  {\rm with}
\beq
        \omega= \rho+\alpha_{3}-\beta_{3},\quad x = \alpha_{1}+\beta_{3}-\alpha_{3}-\rho, \quad
       y= \alpha_{2}+\beta_{3}-\alpha_{3}-\rho,\quad z=\beta_{3}-\rho.
\eeq
where $a_{i}$, $b_{3}$ and all phases are real. Below, in a similar way, we write down the remaining Yukawa textures given in Eq.(\ref{xxx}).
\\
\\
TEXTURE $T_{2}$
\\
\beq
T_{2}=\begin{pmatrix}
a_{1}e^{i\alpha_{1}} & 0\\
a_{2}e^{i\alpha_{2}} & b_{2}e^{i\beta_{2}}\\
a_{3}e^{i\alpha_{3}} & 0
\end{pmatrix}
=
\begin{pmatrix}
e^{ix} & 0&0\\
0 & e^{iy}&0\\
0 & 0&e^{iz}
\end{pmatrix}
\begin{pmatrix}
a_{1} & 0\\
a_{2} & b_{2}\\
a_{3} &0
\end{pmatrix}
\begin{pmatrix}
e^{i\omega} & 0\\
0 & e^{i\rho}
\end{pmatrix},
\eeq {\rm with}
\beq
\omega= \rho+\alpha_{2}-\beta_{2},\quad x= \alpha_{1}+\beta_{2}-\alpha_{2}-\rho,\quad
y= \beta_{2}-\rho,\quad z= \alpha_{3}+\beta_{2}-\alpha_{2}-\rho.
\eeq
TEXTURE $T_{3}$
\\
\beq
T_{3}=\begin{pmatrix}
a_{1}e^{i\alpha_{1}} &b_{1}e^{i\beta_{1}}\\
a_{2}e^{i\alpha_{2}}& 0\\
a_{3}e^{i\alpha_{3}} &0
\end{pmatrix}
=
\begin{pmatrix}
e^{ix} & 0&0\\
0 & e^{iy}&0\\
0 & 0&e^{iz}
\end{pmatrix}
\begin{pmatrix}
a_{1} & b_{1}\\
a_{2} & 0\\
a_{3} &0
\end{pmatrix}
\begin{pmatrix}
e^{i\omega} & 0\\
0 & e^{i\rho}
\end{pmatrix},
\eeq {\rm with}
\beq
\omega= \rho+\alpha_{1}-\beta_{1},\quad x= \beta_{1}-\rho,\quad
y= \alpha_{2}-\alpha_{1}+\beta_{1}-\rho,\quad z=
\alpha_{3}-\alpha_{1}+\beta_{1}-\rho.
\eeq
TEXTURE $T_{4}$
\\
\beq
T_{4}=\begin{pmatrix}
0 & 0\\
a_{2}e^{i\alpha_{2}} & b_{2}e^{i\beta_{2}}\\
a_{3}e^{i\alpha_{3}} &  b_{3}e^{i\beta_{3}}

\end{pmatrix}
=
\begin{pmatrix}
e^{ix} & 0&0\\
0 & e^{iy}&0\\
0 & 0&e^{iz}
\end{pmatrix}
\begin{pmatrix}
0 & 0\\
a_{2} & b_{2}\\
a_{3} &b_{3}e^{i\phi}
\end{pmatrix}
\begin{pmatrix}
e^{i\omega} & 0\\
0 & e^{i\rho}
\end{pmatrix}, \la{t4212}
\eeq
{\rm with}
\beq
\omega=\alpha_{2}-\beta_{2}+\rho,\quad y= \beta_{2}-\rho,\quad z= \alpha_{3}-\alpha_{2}+\beta_{2}-\rho, \quad \phi=\alpha_{2}-\alpha_{3}+\beta_{3}-\beta_{2}. \la{t42121}
\eeq
TEXTURE $T_{5}$
\\
\beq
T_{5}=\begin{pmatrix}
a_{1}e^{i\alpha_{1}} & 0\\
0 & b_{2}e^{i\beta_{2}}\\
a_{3}e^{i\alpha_{3}} & b_{3}e^{i\beta_{3}}

\end{pmatrix}
=
\begin{pmatrix}
e^{ix} & 0&0\\
0 & e^{iy}&0\\
0 & 0&e^{iz}
\end{pmatrix}
\begin{pmatrix}
a_{1} & 0\\
0 & b_{2}\\
a_{3} &b_{3}
\end{pmatrix}
\begin{pmatrix}
e^{i\omega} & 0\\
0 & e^{i\rho}
\end{pmatrix}, \la{t5214}
\eeq {\rm with}
\beq
\omega= \rho+\alpha_{3}-\beta_{3},\quad x= \alpha_{1}+\beta_{3}-\alpha_{3}-\rho,\quad
y= \beta_{2}-\rho,\quad z= \beta_{3}-\rho.
\eeq
TEXTURE $T_{6}$
\\
\beq
T_{6}=\begin{pmatrix}
a_{1}e^{i\alpha_{1}} & 0\\
a_{2}e^{i\alpha_{2}}& b_{2}e^{i\beta_{2}}\\
0 & b_{3}e^{i\beta_{3}}
\end{pmatrix}
=
\begin{pmatrix}
e^{ix} & 0&0\\
0 & e^{iy}&0\\
0 & 0&e^{iz}
\end{pmatrix}
\begin{pmatrix}
a_{1} & 0\\
a_{2} & b_{2}\\
0 &b_{3}
\end{pmatrix}
\begin{pmatrix}
e^{i\omega} & 0\\
0 & e^{i\rho}
\end{pmatrix},
\eeq {\rm with}
\beq
\omega= \rho+\alpha_{2}-\beta_{2}, \quad x= \alpha_{1}+\beta_{2}-\alpha_{2}-\rho, \quad
y= \beta_{2}-\rho, \quad z= \beta_{3}-\rho.
\eeq
TEXTURE $T_{7}$
\\
\beq
T_{7}=\begin{pmatrix}
a_{1}e^{i\alpha_{1}} & b_{1}e^{i\beta_{1}}\\
0 & 0\\
a_{3}e^{i\alpha_{3}} &  b_{3}e^{i\beta_{3}}
\end{pmatrix}
=
\begin{pmatrix}
e^{ix} & 0&0\\
0 & e^{iy}&0\\
0 & 0&e^{iz}
\end{pmatrix}
\begin{pmatrix}
a_{1} & b_{1}\\
0 & 0\\
a_{3} &b_{3}e^{i\phi}
\end{pmatrix}
\begin{pmatrix}
e^{i\omega} & 0\\
0 & e^{i\rho}
\end{pmatrix}, \la{t7218}
\eeq
{\rm with}
\beq
\omega= \rho+\alpha_{1}-\beta_{1}, \quad x= \beta_{1}-\rho, \quad
z= \alpha_{3}-\alpha_{1}+\beta_{1}-\rho,\quad \phi=
\alpha_{1}-\alpha_{3}-\beta_{1}+\beta_{3}.
\eeq
TEXTURE $T_{8}$
\\
\beq
T_{8}=\begin{pmatrix}
a_{1}e^{i\alpha_{1}} &b_{1}e^{i\beta_{1}}\\
a_{2}e^{i\alpha_{2}}& 0\\
0 &b_{3}e^{i\beta_{3}}
\end{pmatrix}
=
\begin{pmatrix}
e^{ix} & 0&0\\
0 & e^{iy}&0\\
0 & 0&e^{iz}
\end{pmatrix}
\begin{pmatrix}
a_{1} & b_{1}\\
a_{2} & 0\\
0 &b_{3}
\end{pmatrix}
\begin{pmatrix}
e^{i\omega} & 0\\
0 & e^{i\rho}
\end{pmatrix},
\eeq {\rm with}
\beq
\omega= \rho+\alpha_{1}-\beta_{1}, \quad x= \beta_{1}-\rho, \quad
y=\alpha_{2}-\alpha_{1}+\beta_{1}-\rho, \quad z= \beta_{3}-\rho.
\eeq
TEXTURE $T_{9}$
\\
\beq
T_{9}=\begin{pmatrix}
a_{1}e^{i\alpha_{1}} & b_{1}e^{i\beta_{1}}\\
a_{2}e^{i\alpha_{2}} & b_{2}e^{i\beta_{2}}\\
0 &  0
\end{pmatrix}
=
\begin{pmatrix}
e^{ix} & 0&0\\
0 & e^{iy}&0\\
0 & 0&e^{iz}
\end{pmatrix}
\begin{pmatrix}
a_{1} & b_{1}\\
a_{2} &  b_{2}e^{i\phi}\\
0 &0
\end{pmatrix}
\begin{pmatrix}
e^{i\omega} & 0\\
0 & e^{i\rho}
\end{pmatrix}, \la{t9222}
\eeq
{\rm with}
\beq
\omega=\alpha_{1}-\beta_{1}+\rho, \quad x= \beta_{1}-\rho, \quad
y= \alpha_{2}-\alpha_{1}+\beta_{1}-\rho, \quad \phi=
\alpha_{1}-\beta_{1}-\alpha_{2}+\beta_{2}. \la{t9223}
\eeq
The phases $x, y$ and $z$ can be eliminated  by proper redefinition of
the states $l$ and $e^c$.
 As far as the phases  $\omega $ and $\rho $ are concerned,
because of the form of the
$M_N$ matrix (\ref{m01}), also they
 will turn out to be non-physical. This is the one main difference of our construction from the scenarios considered earlier \cite{Harigaya:2012bw}.
As we see, in textures $T_{4}$, $T_{7}$ and $T_{9}$ there remains
one unremovable phase $\phi$  (i.e. in the second matrices of the r.h.s of Eqs. (\ref{t4212})
(\ref{t7218}) and (\ref{t9222}) respectively).  This physical phase $\phi$ is relevant to the leptogenesis\cite{Babu:2008kp} and also, as we will see below, it will be related to phase $\delta$, determined from the neutrino sector.
\section{Neutrino mass matrices derived from $2T_0Y_{32}$'s and one ${\rm d}=5$ operator}
\la{derivation}
\numberwithin{equation}{section}
Integrating the RHN's, from the superpotential couplings of Eq. (\ref{r21}), using the see-saw formula, we get the following contribution to the light neutrino mass matrix: \beq M^{ss}_{\nu}=\langle
h^{0}_{u}\rangle^{2} Y_{\nu}M^{-1}_{N}Y^{T}_{\nu}. \la{seesaw} \eeq For $Y_{\nu }$ in (\ref{seesaw}) the textures $T_i$
listed in the previous section should be used in turn. All obtained matrices
$M_{\nu }^{ss}$, if identified with light neutrino mass matrices, will give experimentally unacceptable results. The reason is the number of texture zeros which we have in $T_{i}$ and $M_{N}$ matrices. In order to overcome this difficulty we include the following $\rm d=5$ operator:
 \beq
{\cal
O}^{5}_{ij}\equiv\frac{\tilde{d_{5}}e^{i{x_{5}}}}{2M_{*}}l_{i}l_{j}h_{u}h_{u}
\la{d5}
\eeq where $\tilde{d_{5}}$, $x_{5}$ and $M_{*}$ are real parameters. For each case, we will include a single term of the type of Eq. (\ref{d5}). The latter, together with (\ref{seesaw}) will contribute to the neutrino mass matrix. This will allow to have viable models and, at the same time because of the minimal number of the additions, we will still have predictive scenarios. The operators (\ref{d5}) can be obtained by another sector in such a way as to not affect the forms of $T_{i}$ and $M_{N}$ matrices. We comment about this in Sect. \ref{origin}. Here, we just consider
operators (\ref{d5}) without specifying their origin and
investigate their implications. Recall that, in the previous section, we have written the Yukawa textures in the form: \beq
Y_{\nu}=P_{1}Y^{R}_{\nu}P_{2}, \eeq where $P_{1}, P_{2}$ are diagonal phase
matrices  and $Y^{R}_{\nu}$ is either a real matrix or contains
only one phase (for $T_{4}$, $T_{7}$ and $T_{9}$). Making the field
phase redefinitions:
\beq l^{\prime}=P_{1}l, \quad N^{\prime}=P_{2}N, \quad (e^{\prime})^{c}=P^{*}_{1}e^{c} \eeq
with:
\beq
P_{1}=
\begin{pmatrix}
e^{ix} & 0&0\\
0 & e^{iy}&0\\
0&0&e^{iz}
\end{pmatrix}, \quad P_{2}=
\begin{pmatrix}
e^{i\omega} & 0\\
0 & e^{i\rho}
\end{pmatrix}
\eeq
 the superpotential coupling will become:
 \begin{equation}
W_{e}=(l^{\prime})^{T}Y_{e}^{\rm diag}(e^{\prime})^{c}h_{d},\quad
W_{\nu}=(l^{\prime})^{T}Y^{R}_{\nu}N^{\prime}h_{u}-\frac{1}{2}(N^{\prime})^{T}M^{\prime}_{N}N^{\prime}
\end{equation}
with:
\beq M^{\prime}_{N}=
\begin{pmatrix}
0 & 1\\
1 & 0
\end{pmatrix}\tilde{M}, \quad \tilde{M}=e^{-i(\omega + \rho)}M.
\eeq
Now, for simplification of the notations, we will get rid of the primes (i.e. perform $l^{\prime}\rightarrow l$, $e^{c \prime}\rightarrow e^{c}$,...) and in Eq. (\ref{seesaw}) using $Y_{\nu}^R$ instead of $Y_{\nu}$, from different $T_{i}$ textures we get corresponding $M_{\nu}^{ss}$, and then adding the operator (\ref{d5}), obtain the final neutrino mass matrix.
\\
\\
From textures $T_{1,2,3}$ we obtain:
\beq
M_{T_{1}}=
\begin{pmatrix}
0 & 0&a_{1}b_{3}\\
0 & 0&a_{2}b_{3}\\
a_{1}b_{3}&a_{2}b_{3}&2a_{3}b_{3}
\end{pmatrix}\bar m, \quad
M_{T_{2}}=
\begin{pmatrix}
0 & a_{1}b_{2}&0\\
a_{1}b_{2}&2a_{2}b_{2}&a_{3}b_{2}\\
0&a_{3}b_{2}&0
\end{pmatrix}\bar m, \quad
M_{T_{3}}=
\begin{pmatrix}
2a_{1}b_{1} & a_{2}b_{1}&a_{3}b_{1}\\
a_{2}b_{1}& 0&0\\
a_{3}b_{1}&0&0
\end{pmatrix}\bar m,
\eeq
where $\bar m=\langle
h^{0}_{u}\rangle^{2}/\tilde M$. It is easy to verify that adding one $\rm d=5$ operator mass term to any entry of these mass matrices will not make them experimentally acceptable. Thus, discarding them we move to the remaining textures.
\\
\\
From texture ${T_{4}}$:
\beq
M_{T_{4}}=
\begin{pmatrix}
0 &  0&0\\
0&2a_{2}b_{2}&a_{3}b_{2}+a_{2}b_{3}e^{i\phi}\\
0&a_{3}b_{2}+a_{2}b_{3}e^{i\phi}&2a_{3}b_{3}e^{i\phi}
\end{pmatrix}\bar m.
\eeq
\\
Adding the $\rm d=5$ operators to zero entries of this matrix, we will get three different neutrino mass matrices. Therefore, addition of (\ref{d5}) type term will be performed in the (1,1), (1,2) and (1,3) entries respectively. Since the phase $x$ in Eqs. (\ref{t4212}), (\ref{t42121}) is undetermined, we can shift the phase of state $l_{1}$ in such a way as to match the phase of the (\ref{d5}) operator with the phase of $\bar m$. Thus, this addition will not introduce additional phases inside the neutrino mass matrices. They will have forms:
\\
\beq
M^{(11)}_{T_{4}}=
\begin{pmatrix}
d_{5}& 0&0\\
0&2a_{2}b_{2}&a_{3}b_{2}+a_{2}b_{3}e^{i\phi}\\
0&a_{3}b_{2}+a_{2}b_{3}e^{i\phi}&2a_{3}b_{3}e^{i\phi}
\end{pmatrix}\bar m,
\eeq \beq
M^{(12)}_{T_{4}}=
\begin{pmatrix}
0 & d_{5}&0\\
d_{5}&2a_{2}b_{2}&a_{3}b_{2}+a_{2}b_{3}e^{i\phi}\\
0&a_{3}b_{2}+a_{2}b_{3}e^{i\phi}&2a_{3}b_{3}e^{i\phi}
\end{pmatrix}\bar m, \la{t124}
\eeq \beq
M^{(13)}_{T_{4}}=
\begin{pmatrix}
0 & 0&d_{5}\\
0&2a_{2}b_{2}&a_{3}b_{2}+a_{2}b_{3}e^{i\phi}\\
d_{5}&a_{3}b_{2}+a_{2}b_{3}e^{i\phi}&2a_{3}b_{3}e^{i\phi}
\end{pmatrix}\bar m, \la{t134}
\eeq
where $d_5$ is a real parameter: $d_5=-\tilde d_5 \tilde M/M_*$. By similar way, we will get the other neutrino mass matrices using the remaining Yukawa textures. Also, one can make sure that for those remaining cases there are undetermined phases [see Eqs: (\ref{t5214})-(\ref{t9223})] and their proper shift can match the phase of the term (\ref{d5}) with $\bar m$. Therefore, below, without loss of any generality we can take the parameter $d_5$ (in the neutrino mass matrices) to be real.
\\
\\
From texture ${T_{5}}$:
\beq
M_{T_{5}}=
\begin{pmatrix}
0 & a_{1}b_{2}&a_{1}b_{3}\\
a_{1}b_{2} & 0&a_{3}b_{2}\\
a_{1}b_{3}&a_{3}b_{2}&2a_{3}b_{3}
\end{pmatrix}\bar m.
\eeq
\beq
M^{(11)}_{T_{5}}= \left(\begin{array}{ccc}
 d_{5} & a_{1}b_{2}&a_{1}b_{3}\\
a_{1}b_{2} & 0&a_{3}b_{2}\\
a_{1}b_{3}&a_{3}b_{2}&2a_{3}b_{3}
\end{array}\right)\bar m, \quad
M^{(22)}_{T_{5}}= \left(\begin{array}{ccc}
 0 & a_{1}b_{2}&a_{1}b_{3}\\
a_{1}b_{2} & d_{5}&a_{3}b_{2}\\
a_{1}b_{3}&a_{3}b_{2}&2a_{3}b_{3}
\end{array}\right)\bar m.
\eeq
From texture ${T_{6}}$:
\beq
M_{T_{6}}=
\begin{pmatrix}
0 & a_{1}b_{2}&a_{1}b_{3}\\
a_{1}b_{2} &2a_{2}b_{2}&a_{2}b_{3}\\
a_{1}b_{3}&a_{2}b_{3}&0
\end{pmatrix}\bar m.
\eeq
\beq
M^{(33)}_{T_{6}}= \left(\begin{array}{ccc}
 0 & a_{1}b_{2}&a_{1}b_{3}\\
a_{1}b_{2} &2a_{2}b_{2}&a_{2}b_{3}\\
a_{1}b_{3}&a_{2}b_{3}&d_{5}
\end{array}\right)\bar m, \quad
M^{(11)}_{T_{6}}= \left(\begin{array}{ccc}
 d_{5}& a_{1}b_{2}&a_{1}b_{3}\\
a_{1}b_{2} &2a_{2}b_{2}&a_{2}b_{3}\\
a_{1}b_{3}&a_{2}b_{3}&0
\end{array}\right)\bar m.
\eeq
From texture ${T_{7}}$:
\beq
M_{T_{7}}=
\begin{pmatrix}
2a_{1}b_{1} &  0&a_{3}b_{1}+a_{1}b_{3}e^{i\phi}\\
0&0&0\\
a_{3}b_{1}+a_{1}b_{3}e^{i\phi}&0&2a_{3}b_{3}e^{i\phi}
\end{pmatrix}\bar m.
\eeq
\beq
M^{(22)}_{T_{7}}= \begin{pmatrix}
2a_{1}b_{1} &  0&a_{3}b_{1}+a_{1}b_{3}e^{i\phi}\\
0&d_{5}&0\\
a_{3}b_{1}+a_{1}b_{3}e^{i\phi}&0&2a_{3}b_{3}e^{i\phi}
\end{pmatrix}\bar m,
\eeq \beq
M^{(12)}_{T_{7}}=\begin{pmatrix}
2a_{1}b_{1} &  d_{5}&a_{3}b_{1}+a_{1}b_{3}e^{i\phi}\\
d_{5}&0&0\\
a_{3}b_{1}+a_{1}b_{3}e^{i\phi}&0&2a_{3}b_{3}e^{i\phi}
\end{pmatrix}\bar m,
\eeq \beq
M^{(23)}_{T_{7}}=
\begin{pmatrix}
2a_{1}b_{1} &  0&a_{3}b_{1}+a_{1}b_{3}e^{i\phi}\\
0&0&d_{5}\\
a_{3}b_{1}+a_{1}b_{3}e^{i\phi}&d_{5}&2a_{3}b_{3}e^{i\phi}
\end{pmatrix}\bar m.\la{t237}
\eeq
From texture ${T_{8}}$:
\beq
M_{T_{8}}=
\begin{pmatrix}
2a_{1}b_{1} & a_{2}b_{1}&a_{1}b_{3}\\
a_{2}b_{1}& 0&a_{2}b_{3}\\
a_{1}b_{3}&a_{2}b_{3}&0
\end{pmatrix}\bar m.
\eeq
\beq
M^{(22)}_{T_{8}}= \left(\begin{array}{ccc}
2a_{1}b_{1} & a_{2}b_{1}&a_{1}b_{3}\\
a_{2}b_{1}& d_{5}&a_{2}b_{3}\\
a_{1}b_{3}&a_{2}b_{3}&0
\end{array}\right)\bar m, \quad
M^{(33)}_{T_{8}}= \left(\begin{array}{ccc}
2a_{1}b_{1} & a_{2}b_{1}&a_{1}b_{3}\\
a_{2}b_{1}& 0&a_{2}b_{3}\\
a_{1}b_{3}&a_{2}b_{3}&d_{5}
\end{array}\right)\bar m.
\eeq
From texture ${T_{9}}$:
\beq
M_{T_{9}}=
\begin{pmatrix}
2a_{1}b_{1} & a_{2}b_{1}+a_{1}b_{2}e^{i\phi}&0\\
a_{2}b_{1}+a_{1}b_{2}e^{i\phi}&2a_{2}b_{2}e^{i\phi} &0\\
0&0&0
\end{pmatrix}\bar m.
\eeq
\beq
M^{(13)}_{T_{9}}= \begin{pmatrix}
2a_{1}b_{1} & a_{2}b_{1}+a_{1}b_{2}e^{i\phi}&d_{5}\\
a_{2}b_{1}+a_{1}b_{2}e^{i\phi}&2a_{2}b_{2}e^{i\phi} &0\\
d_{5}&0&0
\end{pmatrix}\bar m,
\eeq \beq
M^{(23)}_{T_{9}}=
\begin{pmatrix}
2a_{1}b_{1} & a_{2}b_{1}+a_{1}b_{2}e^{i\phi}&0\\
a_{2}b_{1}+a_{1}b_{2}e^{i\phi}&2a_{2}b_{2}e^{i\phi} &d_{5}\\
0&d_{5}&0
\end{pmatrix}\bar m, \la{t239}
\eeq \beq
M^{(33)}_{T_{9}}= \begin{pmatrix}
2a_{1}b_{1} & a_{2}b_{1}+a_{1}b_{2}e^{i\phi}&0\\
a_{2}b_{1}+a_{1}b_{2}e^{i\phi}&2a_{2}b_{2}e^{i\phi} &0\\
0&0&d_{5}
\end{pmatrix}\bar m.
\eeq
We have shown that only $T_{4}$, $T_{7}$ and $T_{9}$
$2T_0Y_{32}$'s give rise to complex mass matrices and  that complexity, i.e.  phase $\delta $ in the lepton mixing matrix, arises through (\ref{seesaw}) $\text{---}$ from complex $2T_0Y_{32}$'s $\text{---}$ and not from an $x_{5}$ phase.
\subsection{Possible origin of $\rm d=5$ operators}
\la{origin}
The $\rm d=5$ operator coupling  [see Eq. (\ref{d5})] in our case has been directly introduced in the neutrino mass matrices.
Here we give one example of possible generation of $\rm d=5$ operators  we are exploiting within our setup.
Besides being of a quantum gravity origin, such $\rm d=5$ couplings can be generated from a different sector via renormalizable interactions.
For instance, introducing the pair of MSSM singlet states ${\cal N}$, $\overline{\cal N}$ and the superpotential couplings
\begin{equation}
\lam^{(i)}l_i{\cal N}h_u+\bar \lam^{(j)}l_j\overline{\cal N}h_u-M_*{\cal N}\overline{\cal N}~,
\end{equation}
it is easy to verify that integration of the heavy ${\cal N}$, $\overline{\cal N}$ multiplets leads to the operator in Eq. (\ref{d5})
with
\begin{equation}
\tilde {d_5}e^{ix_5}=2\lam^{(i)}\bar \lam^{(j)}~.
\end{equation}
Important ingredient here is to maintain forms of the resulting mass matrices and do not mix the states  ${\cal N}$, $\overline{\cal N}$
with RHN's $N_{1,2}$. This can be achieved by some (possible flavor) symmetries (which we do not pursue here).
Perhaps a safer way to generate those $\Delta L=2$ effective couplings would be to proceed in a spirit of type II \cite{{Magg:1980ut},{Lazarides:1980nt},{Mohapatra:1980yp}}, or  type III   \cite{Foot:1988aq},\cite{Ma:1998dn} see-saw
mechanisms, or exploit alternative possibilities \cite{{Zee:1980ai},{Babu:1988ki},{Babu:2011vb},{Tavartkiladze:2001by},{Perez:2008ha},{Bajc:2007zf},{Babu:2009aq},{Bonnet:2009ej},{Kumericki:2012bh},{Xing:2009hx},{Gavela:2009cd}}. through the introduction of appropriate extra states. Details of such scenarios should be pursued elsewhere.
\section{Analyzing neutrino mass matrices}
\la{analysis}
\numberwithin{equation}{section}
Since we are working in a basis of a diagonal charged lepton mass matrix, lepton mixing matrix $U$  entirely comes from the neutrino sector. Therefore, the following equality holds: \beq
M_{\nu}=PU^{*}P^{'}M_{\nu}^{\rm diag}U^{+}P \la{nu1} \eeq where
\beq
M_{\nu}^{\rm diag}=(m_{1},m_{2},m_{3}), \quad P={\rm
Diag}(e^{i\omega_{1}},e^{i\omega_{2}},e^{i\omega_{3}}),\quad
P^{'}={\rm Diag}(1,e^{i\rho_{1}},e^{i\rho_{2}})\la{nu2}
\eeq \beq
U= \left(\begin{array}{ccc}
c_{13}c_{12} &c_{13}s_{12}&s_{13}e^{-i\delta}\\
-c_{23}s_{12}-s_{23}s_{13}c_{12}e^{i\delta}& c_{23}c_{12}-s_{23}s_{13}s_{12}e^{i\delta}&s_{23}c_{13}\\
s_{23}s_{12}-c_{23}s_{13}c_{12}e^{i\delta}&-s_{23}c_{12}-c_{23}s_{13}s_{12}e^{i\delta}&c_{23}c_{13}
\end{array}\right)\la{nu3}
\eeq
where $m_{i}$ denote neutrino masses. $U$ given in Eq. (\ref{nu3}) is the standard parametrization used in the literature (see for instance \cite{Fogli:2012ua}, \cite{Agashe:2014kda}). The relation (\ref{nu1}) turns out to be convenient and useful for neutrino mass matrix analysis. Numerical values of oscillation parameters both, for normal (NH) and inverted (IH) hierarchies can be found in \cite{Gonzalez-Garcia:2014bfa}. Thus, for these mass orderings we will use the following notations:
\\
\begin{center}
{\centering For normal hierarchy (NH):}
\end{center}
\beq \Delta
m_{sol}^{2}=m_{2}^{2}-m_{1}^{2},\quad
\Delta
m_{atm}^{2}=m_{3}^{2}-m_{2}^{2},\quad
m_{1}=\sqrt{m_{3}^{2}-\Delta m_{atm}^{2}-\Delta m_{sol}^{2}},\quad
m_{2}=\sqrt{m_{3}^{2}-\Delta m_{atm}^{2}} \la{nh1} \eeq
\begin{center}
{\centering For inverted hierarchy (IH)}
\end{center}
\beq \Delta
m_{atm}^{2}=m_{2}^{2}-m_{3}^{2},\quad
\Delta
m_{sol}^{2}=m_{2}^{2}-m_{1}^{2},\quad
m_{1}=\sqrt{m_{3}^{2}+\Delta m_{atm}^{2}-\Delta m_{sol}^{2}},\quad
m_{2}=\sqrt{m_{3}^{2}+\Delta m_{atm}^{2}} \la{ih1}
\eeq
\subsection{Types of neutrino mass matrices}
Complex $3\times3$ Majorana type neutrino mass matrices with more than two independent zero entries are all excluded by current experiments. As it turns out, experimental data also exclude the possibility of real neutrino mass
matrices with two independent zero entries. This was noticed earlier upon studies of the texture zero neutrino mass matrices \cite{Frampton:2002qc},\cite{Frampton:2002yf},\cite{Fritzsch:2011qv},\cite{Dev:2006qe}. Therefore,  experimentally viable neutrino mass matrices, from our $3\times2$ Yukawa textures (listed in Sect. \ref{matrices}) should be produced by  $T_4,..., T_9$ giving either neutrino mass matrices with two independent zero entries and the complex phase, or the one zero entry real neutrino mass matrices (via textures $T_{5}$, $T_{6}$, $T_{8}$ and one d=5 operator).
Two zero entry complex neutrino mass matrices (we have obtained) have forms:
\beq
P_{1}=\left(\begin{array}{ccc}
0 & \times&0\\
\times& \times&\times\\
0&\times&\times
\end{array}\right),\quad
P_{2}=\left(\begin{array}{ccc}
0& 0&\times\\
0&\times&\times\\
\times&\times&\times
\end{array}\right),\quad
P_{3}=\left(\begin{array}{ccc}
\times &0&\times\\
0& 0&\times\\
\times&\times&\times
\end{array}\right),\quad
P_{4}=\left(\begin{array}{ccc}
\times& \times&0\\
\times& \times&\times\\
0&\times&0
\end{array}\right). \la{pse}
\eeq
These types of textures correspond to the following mass matrices, we have obtained:
\\
\\
$P_{1}$-type:\quad$M^{(12)}_{T_{4}}$,\qquad$P_{2}$-type:\quad$M^{(13)}_{T_{4}}$,\quad$P_{3}$-type:\quad$M^{(23)}_{T_{7}}$,\qquad$P_{4}$-type:\quad$M^{(23)}_{T_{9}}$
\\
\\
As far as the one zero entry neutrino mass matrices are concerned we are getting the following types of real mass matrices:
\beq
P_{5}=\left(\begin{array}{ccc}
0 & \times&\times\\
\times& \times&\times\\
\times&\times&\times
\end{array}\right),\quad
P_{6}=\left(\begin{array}{ccc}
\times&\times&\times\\
\times&0&\times\\
\times&\times&\times
\end{array}\right),\quad
P_{7}=\left(\begin{array}{ccc}
\times &\times&\times\\
\times& \times&\times\\
\times&\times&0
\end{array}\right). \la{ps1}
\eeq
Also here, we indicate the correspondence of $P_{5,6,7}$ textures to the appropriate neutrino mass matrices we have obtained: $P_{5}$-type:\quad$M^{(22)}_{T_{5}}$,\quad$M^{(33)}_{T_{6}}$,\qquad$P_{6}$-type:\quad$M^{(11)}_{T_{5}}$,\quad$M^{(33)}_{T_{8}}$ and\qquad$P_{7}$-type:\quad$M^{(11)}_{T_{6}}$,\quad$M^{(22)}_{T_{8}}$.
\subsection{Predictions from $P_{1,2,3,4}$ type neutrino mass matrices}
Here we analyze neutrino mass matrices with two independent zero entries.
As we will see, for each case we will get several predictions.
\\
\\
TYPE $P_{1}$
\\
Structure of the $P_{1}$ in Eq.(\ref{pse}) imposes the following conditions: $M^{(1,1)}_{\nu}=0$ and $M^{(1,3)}_{\nu}$=0, which taking into account (\ref{nu1})-(\ref{nu3}) give the following relations:  \beq
\frac{m_{1}}{m_{3}}c^{2}_{12}+\frac{m_{2}}{m_{3}}s^{2}_{12}e^{i\rho_{1}}=-t^{2}_{13}e^{i(\rho_{2}+2\delta)}
\la{b} \eeq and \beq
-\left(\frac{m_{1}}{m_{3}}-\frac{m_{2}}{m_{3}}e^{i\rho_{1}}\right)t_{23}s_{12}c_{12}-s_{13}e^{i(\rho_{2}+\delta)}+s_{13}e^{-i\delta}\left(\frac{m_{1}}{m_{3}}c^{2}_{12}+\frac{m_{2}}{m_{3}}s^{2}_{12}e^{i\rho_{1}}\right)=0\quad
\la{c} \eeq Using  (\ref{b}) in the last term of (\ref{c}) we
obtain: \beq
\left(\frac{m_{1}}{m_{3}}-\frac{m_{2}}{m_{3}}e^{i\rho_{1}}\right)t_{23}s_{12}c_{12}+s_{13}e^{i(\rho_{2}+\delta)}+s_{13}t^{2}_{13}e^{i(\rho_{2}+\delta)}=0
\la{d}
\eeq which gives: \beq
m_{3}s_{13}(1+t^{2}_{13})=|m_{1}-m_{2}e^{i\rho_{1}}|t_{23}s_{12}c_{12}
\la{e} \eeq while from Eq. (\ref{b}) we have: \beq
m_{3}t^{2}_{13}=|m_{1}c^{2}_{12}+m_{2}s^{2}_{12}e^{i\rho_{1}}|.
\la{f}
\eeq We can exclude phase $\rho_{1}$ from (\ref{e}) and (\ref{f})
to obtain: \beq
m^{2}_{3}(t^{4}_{13}+s^{2}_{13}\cot^{2}_{23}(1+t^{2}_{13})^{2})=m^{2}_{1}c^{2}_{12}+m^{2}_{2}s^{2}_{12}
 \la{g}
\eeq
From which, based on recent experimental data \cite{Gonzalez-Garcia:2014bfa} inverted hierarchical pattern (IH) is excluded. For normal hierarchical neutrinos from (\ref{g}), with
(\ref{nh1}) we get \beq
m^{2}_{3}=\frac{\Delta m^{2}_{atm}+\Delta
m^{2}_{sol}c^{2}_{12}}{1-s^{2}_{13}\cot^{2}_{23}(1+t^{2}_{13})^{2}-t^{4}_{13}}.
\la{h} \eeq
Using $\sin^2\theta_{23}=0.49$, the best fit values for the remaining mixing angles \cite{Gonzalez-Garcia:2014bfa} and also the best fit values for the atmospheric and solar neutrino mass squared differences:\beq
\Delta m^{2}_{atm}=0.002382 ~\rm eV^{2},\quad \Delta
m^{2}_{sol}=7.5\times10^{-5} ~\rm eV^{2} \la{i}
\eeq
from (\ref{h}) we obtain for NH: \beq
m_{1}=0.00613 ~\rm eV,\quad m_{2}=0.0106 ~\rm eV,\quad m_{3}=0.0499 ~\rm eV.
\la{k}
\eeq Using these, from (\ref{f}) we predict: \beq
\cos\rho_{1}=\frac{m^{2}_{3}t^{4}_{13}-m^{2}_{1}c^{4}_{12}-m^{2}_{2}s^{4}_{12}}{2m_{1}m_{2}c^{2}_{12}s^{2}_{12}}
\Rightarrow \rho_{1}=\pm 3.036, \la{l}
\eeq while from (\ref{b}) and (\ref{d}) we have: \beqs
\delta=\arg[m_{1}c^{2}_{12}+m_{2}s^{2}_{12}e^{i\rho_{1}}]-\arg[m_{1}-m_{2}e^{i\rho_{1}}],
\eeqs \beq
\rho_{2}=\pm
\pi-\arg[m_{1}c^{2}_{12}+m_{2}s^{2}_{12}e^{i\rho_{1}}]+2\arg[m_{1}-m_{2}e^{i\rho_{1}}].
\la{m}
\eeq With numbers given in (\ref{k}) and (\ref{l}),
from (\ref{m}) we obtain: \beq
\delta =\pm 0.378,\quad \rho_1=\pm 3.036,\quad \rho_2=\pm 2.696,\quad m_{\beta\beta}=0,
\eeq
where the neutrino-less double beta decay parameter $m_{\beta\beta }$ is
determined as:
$m_{\beta\beta}=|m_{1}c_{12}^{2}c_{13}^{2}+m_{2}e^{i\rho_{1}}c_{13}^{2}s_{12}^{2}+m_{3}e^{i\rho_{2}}s_{13}^{2}e^{2i\delta}|$.
We summarize our results in Table \ref{tab1}.
\begin{center}
  \begin{tabular}{|l|r|r|r|}
  \hline
  \multicolumn{1}{|c|}{\sffamily $\delta$}
 &\multicolumn{1}{|c|}{\sffamily $\rho_{1}$}&\multicolumn{1}{|c|}{\sffamily $\rho_{2}$}&\multicolumn{1}{|c|}{\sffamily works with}\\
  \hline
  $\delta =\pm 0.378$&$\rho_1=\pm 3.036$&$\rho_2=\pm 2.696$&\makecell{NH, $\sin^{2}\theta_{23}=0.49$ and best\\ fit values for remaining oscillation parameters,\\ $(m_{1},m_{2},m_{3})=(0.00613,0.0106,0.0499)$, $m_{\beta\beta}=0$}\\
  \hline
  \end{tabular}
  \captionof{table}{Results from $P_{1}$ type texture. Masses are given in eVs.}
  \label{tab1}
  \end{center}
TYPE $P_{2}$
\\
In this case $M^{(1,1)}_{\nu}=0$ and $M^{(1,2)}_{\nu}$=0 and together with Eq.(\ref{b}), the following relation holds: \beq
-\left(\frac{m_{1}}{m_{3}}-\frac{m_{2}}{m_{3}}e^{i\rho_{1}}\right)s_{12}c_{12}+s_{13}t_{23}e^{i(\rho_{2}+\delta)}-s_{13}t_{23}e^{-i\delta}\left(\frac{m_{1}}{m_{3}}c^{2}_{12}+\frac{m_{2}}{m_{3}}s^{2}_{12}e^{i\rho_{1}}\right)=0.
\la{b1}
\eeq Using (\ref{b}) in the last term of (\ref{b1}) we obtain:
\beq
-\left(\frac{m_{1}}{m_{3}}-\frac{m_{2}}{m_{3}}e^{i\rho_{1}}\right)s_{12}c_{12}+s_{13}t_{23}e^{i(\rho_{2}+\delta)}+s_{13}t_{23}t^{2}_{13}e^{i(\rho_{2}+\delta)}=0
\la{c1}
\eeq which gives: \beq
m_{3}s_{13}t_{23}(1+t^{2}_{13})=|m_{1}-m_{2}e^{i\rho_{1}}|s_{12}c_{12}.
\la{d1}
\eeq Excluding phase $\rho_{1}$ from Eqs. (\ref{d1}) and (\ref{f})[which is derived from Eq.(\ref{b}),  i.e. the condition $M^{(1,1)}_{\nu}=0$]
we obtain: \beq
m^{2}_{3}(t^{4}_{13}+s^{2}_{13}t^{2}_{23}(1+t^{2}_{13})^2)=m^{2}_{1}c^{2}_{12}+m^{2}_{2}s^{2}_{12}
\la{e1}
\eeq Last relation makes obvious that the IH case is
excluded. On the other hand, for NH neutrinos,
from (\ref{e1}), with (\ref{nh1}) we get: \beq
m^{2}_{3}=\frac{\Delta m^{2}_{atm}+\Delta
m^{2}_{sol}c^{2}_{12}}{1-s^{2}_{13}t^{2}_{23}(1+t^{2}_{13})^{2}-t^{4}_{13}}.\la{f1}
\eeq
After finding the value of $m_3$
and remaining masses,
\beq
(m_{1},m_{2},m_{3})=(0.00501,0.01,0.04982)~\rm eV.
\eeq
Eqs. (\ref{b1}) and (\ref{c1}) allow to calculate the phases:
\beq
\cos\rho_{1}=\frac{m^{2}_{3}t^{4}_{13}-m^{2}_{1}c^{4}_{12}-m^{2}_{2}s^{4}_{12}}{2m_{1}m_{2}c^{2}_{12}s^{2}_{12}}
\Rightarrow \rho_{1}=\mp 2.828,\la{i1}
\eeq\beqs
\delta=\pm\pi+\arg[m_{1}c^{2}_{12}+m_{2}s^{2}_{12}e^{i\rho_{1}}]-\arg[m_{1}-m_{2}e^{i\rho_{1}}],
\eeqs \beq
\rho_{2}=\mp
\pi-\arg[m_{1}c^{2}_{12}+m_{2}s^{2}_{12}e^{i\rho_{1}}]+2\arg[m_{1}-m_{2}e^{i\rho_{1}}].
\la{j1}
\eeq
Using the best fit values of measured parameters
\cite{Gonzalez-Garcia:2014bfa} for NH we obtain results
 \beq
\delta =\pm 1.924,\quad \rho_1=\mp 2.828,\quad \rho_2=\mp 1.715,\quad m_{\beta\beta}=0, \la{n}
\eeq
which are summarized in Table \ref{tab2}:
\begin{center}
  \begin{tabular}{|l|r|r|r|}
  \hline
  \multicolumn{1}{|c|}{\sffamily $\delta$}
 &\multicolumn{1}{|c|}{\sffamily $\rho_{1}$}&\multicolumn{1}{|c|}{\sffamily $\rho_{2}$}&\multicolumn{1}{|c|}{\sffamily works with}\\
  \hline
  $\delta =\pm 1.924$&$\rho_1=\mp 2.828$&$\rho_2=\mp 1.715$&\makecell{NH and best fit\\ values of oscillation parameters,\\ $(m_{1},m_{2},m_{3})=(0.00501,0.01,0.04982)$, $m_{\beta\beta}=0$}\\
  \hline
  \end{tabular}
  \captionof{table}{Results from $P_{2}$ type texture. Masses are given in eVs.}
  \label{tab2}
  \end{center}
$P_1$ and $P_2$ neutrino textures were studied in \cite{{Fritzsch:2011qv},{Dev:2006qe}, {Xing:2002ta},{Grimus:2011sf},{Dev:2014dla},{Zhou:2015qua},{Kitabayashi:2015jdj}}. Our analytical expressions, allowing thorough investigations, are
compact and exact.
To analyze the textures $P_{3}$ and $P_{4}$ it is
convenient to note, that equation $M^{(i,j)}_{\nu}=0$ can be
written as: $A_{2}\times m_{2} e^{i\rho_{1}}+A_{3}\times m_{3}
e^{i\rho_{2}}=A_{1}\times m_{1}$. When two mass matrix elements
are equal to zero we have a pair of similar equations which we
write in a matrix form: \beq
 \left(\begin{array}{ccc}
A_{2} &A_{3}\\
B_{2}& B_{3}
\end{array}\right)
\binom{m_{2} e^{i\rho_{1}}}{m_{3} e^{i\rho_{2}}}=
\binom{A_{1}m_{1}}{B_{1}m_{1}}.\eeq From these equations we have: \beq m_{2}
e^{i\rho_{1}}=\frac{1}{A_{2}B_{3}-A_{3}B_{2}}(B_{3}A_{1}-A_{3}B_{1}) m_{1},
\quad  m_{3}
e^{i\rho_{2}}=\frac{1}{A_{2}B_{3}-A_{3}B_{2}}(A_{2}B_{1}-B_{2}A_{1}) m_{1}\la{p1}\eeq
or, \beq
m^{2}_{2}=\frac{|B_{3}A_{1}-A_{3}B_{1}|^{2}}{|A_{2}B_{3}-A_{3}B_{2}|^{2}}m^{2}_{1},
\quad
m^{2}_{3}=\frac{|A_{2}B_{1}-B_{2}A_{1}|^{2}}{|A_{2}B_{3}-A_{3}B_{2}|^{2}}m^{2}_{1}\la{p2}\eeq
and \beq \frac{\Delta m^{2}_{sol}}{\pm\Delta
m^{2}_{atm}}=\frac{|B_{3}A_{1}-A_{3}B_{1}|^{2}-|A_{2}B_{3}-A_{3}B_{2}|^{2}}{|A_{2}B_{1}-B_{2}A_{1}|^{2}-|B_{3}A_{1}-A_{3}B_{1}|^{2}},
\la{p3} \eeq where "+" and "-" signs correspond to normal and inverted hierarchies respectively.  Eq. (\ref{p3}) is the relation for calculating the value of $\delta $. At the same time
(after knowing the $\delta $), from Eq. (\ref{p2}) and (\ref{nh1})/(\ref{ih1}) the neutrino masses can be calculated. After these, with relations in Eq. (\ref{p1}) the phases $\rho_1$ and $\rho_2$ can be found. Below, we use this procedure for the textures $P_3$ and $P_4$.
\\
\\
TYPE $P_{3}$
\\
For this case we have:
\beqs
A_{1}=-U^{\ast}_{11}U^{\dagger}_{12},\quad
A_{2}=U^{\ast}_{12}U^{\dagger}_{22},\quad
A_{3}=U^{\ast}_{13}U^{\dagger}_{32},\quad
B_{1}=-U^{\ast}_{21}U^{\dagger}_{12},\quad
B_{2}=U^{\ast}_{22}U^{\dagger}_{22},\quad
B_{3}=U^{\ast}_{23}U^{\dagger}_{32}\eeqs
and using these in Eqs. (\ref{p1})-(\ref{p3}), for NH and IH neutrino mass ordering, we get results which are summarized in Table \ref{tab3'}.
\begin{center}
  \begin{tabular}{|l|r|r|r|}
  \hline
  \multicolumn{1}{|c|}{\sffamily $\delta$}
 &\multicolumn{1}{|c|}{\sffamily $\rho_{1}$}&\multicolumn{1}{|c|}{\sffamily $\rho_{2}$}&\multicolumn{1}{|c|}{\sffamily works with}\\
  \hline
 $\delta =\pm 1.547$&$\rho_1=\pm 0.0615$&$\rho_2=\mp 3.098$&\makecell{NH and best fit values\\ of oscillation parameters,\\ $(m_{1},m_{2},m_{3})=$\\$(0.07213,0.07265,0.08752)$,\\ $m_{\beta\beta}=0.0726$}\\
  \hline
  $\delta =\pm 1.579$&$\rho_1=\mp 0.0998$&$\rho_2=\pm 3.0726$&\makecell{IH and best fit values\\ of oscillation parameters,\\ $(m_{1},m_{2},m_{3})=$\\$(0.07195,0.07247,0.05294)$,\\ $m_{\beta\beta}=0.0716$}\\
  \hline
  \end{tabular}
  \captionof{table}{Results from $P_{3}$ type texture. Masses are given in eVs.}
  \label{tab3'}
  \end{center}
TYPE $P_{4}$
\\
For this case we have:\beqs
A_{1}=-U^{\ast}_{11}U^{\dagger}_{13},\quad
A_{2}=U^{\ast}_{12}U^{\dagger}_{23},\quad
A_{3}=U^{\ast}_{13}U^{\dagger}_{33},\quad
B_{1}=-U^{\ast}_{31}U^{\dagger}_{13},\quad
B_{2}=U^{\ast}_{32}U^{\dagger}_{23},\quad
B_{3}=U^{\ast}_{33}U^{\dagger}_{33}.\eeqs
For this case NH works with
$\sin^2 \theta_{23}$ larger by $1\sigma $ from the best fit value. However, IH case requires a lower value of $\sin^2 \theta_{23}$. Using above relations in Eqs. (\ref{p1})-(\ref{p3}), for NH and IH cases we get results which are summarized in Table \ref{tab4}.
\begin{center}
  \begin{tabular}{|l|r|r|r|}
  \hline
  \multicolumn{1}{|c|}{\sffamily $\delta$}
 &\multicolumn{1}{|c|}{\sffamily $\rho_{1}$}&\multicolumn{1}{|c|}{\sffamily $\rho_{2}$}&\multicolumn{1}{|c|}{\sffamily works with}\\
 \hline
 $\delta =\pm 1.575$&$\rho_1=\mp 0.0127$&$\rho_2=\pm 3.133$&\makecell{NH and $\sin^2 \theta_{23}=0.51$ and best fit values\\ for remaining oscillation parameters,\\ $(m_{1},m_{2},m_{3})=$\\$(0.171701,0.171919,0.1787)$,\\ $m_{\beta\beta}=0.1719$}\\
  \hline
  $\delta =\pm 1.5705$&$\rho_1=\pm 0.00622$&$\rho_2=\mp 3.137$&\makecell{IH and $\sin^2 \theta_{23}=0.495$ and best fit values\\ for remaining oscillation parameters,\\ $(m_{1},m_{2},m_{3})=$\\$(0.2513,0.25145,0.2465)$,\\ $m_{\beta\beta}=0.2512$}\\
  \hline
  \end{tabular}
  \captionof{table}{Results from $P_{4}$ type texture. Masses are given in eVs.}
  \label{tab4}
  \end{center}
Our results for the textures $P_3$ and $P_4$ are compatible with ones \cite{{Fritzsch:2011qv},{Dev:2006qe}, {Xing:2002ta},{Grimus:2011sf},{Dev:2014dla},{Zhou:2015qua},{Kitabayashi:2015jdj},{Meloni:2014yea}}, obtained before.\footnote{Some of these works used the earlier experimental data. We have made sure, that with those inputs, we would get similar results.}
\subsection{Predictions from real one zero entry neutrino textures - $P_{5,6,7}$}
Now we turn to the analysis of the one texture zero neutrino mass matrices we have obtained in Section 3.  They fall in the category of the $P_{5,6,7}$ type
mass matrices given in Eq. (\ref{ps1}). One texture zero neutrino mass matrices were investigated in \cite{{Merle:2006du},{Lashin:2011dn},{Deepthi:2011sk},{Liao:2013rca},{Gautam:2015kya}}. In our construction, these mass matrices are real. This makes them more predictive.
\\
\\
TYPE $P_{5}$
\\
In this case, our construction implies $\phi$=0 and all elements of the lepton mixing matrix are real (i.e. $\delta$=0 or $\pi$). Therefore, together with $M_{\nu }^{(1,1)}$=0 we have to match phases of both sides of Eq.(\ref{nu1}). This turns out to be impossible for $\rho_{1},\rho_{2}$ not equal to either  0 or $\pi$ , because we have only three free phases $\omega_{1,2,3}$. Thus, it turns out that only normal hierarchical scenario will be allowed with $\delta =0$ or $\pi $. With these, and from the condition $M_{\nu }^{(1,1)}$=0, we get
\beq
\tan\theta_{13}=\left(-c_1c_2s_{12}^{2}\frac{m_{2}}{m_{3}}-c_2c_{12}^{2}\frac{m_{1}}{m_{3}}\right)^{\frac{1}{2}}, \la{tta}
\eeq
where $c_1$ and $c_2$ stand for
$\cos \rho_1$ and $\cos \rho_2$ respectively.
This relation can be satisfied by special selection of the neutrino masses and $\rho_{1,2}=0$ or $\pi $. Since two mass square differences are fixed from the neutrino data, only one free mass is available, which we choose to be $m_{3}$. The latter is tightly constrained via Eq.(\ref{tta}).
Thus, the model predicts three neutrino masses and the phases. For the best fit values of the oscillation parameters \cite{Gonzalez-Garcia:2014bfa} for NH we obtain solutions:
\\
\beqs
m_{1}=0.002268~\rm eV,\quad    m_{2}=0.008952~\rm eV, \quad    m_{3}=0.04962~\rm eV, \eeqs \beq  {\rm with}\quad m_{\beta\beta}=0, \quad \delta=0 \quad{\rm or}\quad \pi,  \quad \rho_{1}=\pi, \quad \rho_{2}=0
\eeq
\\
and
\beqs
m_{1}=0.010677~\rm eV,\quad    m_{2}=0.006245~\rm eV, \quad    m_{3}=0.04996~\rm eV, \eeqs \beq  {\rm with}\quad m_{\beta\beta}=0, \quad \delta=0\quad{\rm or} \quad\pi,  \quad \rho_{1}=\pi, \quad \rho_{2}=\pi.
\eeq
By the similar analysis, we can easily make sure that inverted hierarchy is not allowed within our construction for this $P_{5}$ type texture.
\\
\\
TYPE $P_{6}$
\\
For this case, the condition
$M_{\nu }^{(2,2)}=0$ gives the
following expression for $\theta_{12}$:
\beqs
\tan \theta_{12}=\frac{c_{23}s_{23}\hat{s}_{13}(m_{2}c_{1}-m_{1})}{m_{1}c_{23}^{2}+m_{2}s_{23}^{2}s_{13}^{2}c_{1}+m_{3}s_{23}^{2}c_{13}^{2}c_{2}}
\eeqs
\beq
\pm\frac{\sqrt{c^2_{23}s^2_{23}s^2_{13}(m_{2}c_{1}-m_{1})^2-(m_{1}c_{23}^{2}+m_{2}s_{23}^{2}s_{13}^{2}c_{1}+m_{3}s_{23}^{2}c_{13}^{2}c_{2})(m_{1}s_{23}^{2}s_{13}^{2}+m_{2}c_{23}^{2}c_{1}+m_{3}s_{23}^{2}c_{13}^{2}c_{2})}}{m_{1}c_{23}^{2}+m_{2}s_{23}^{2}s_{13}^{2}c_{1}+m_{3}s_{23}^{2}c_{13}^{2}c_{2}} \la{tpm1}
\eeq
where, $c_{1}$ and $c_{2}$ stand for $\cos\rho_{1}$ and $\cos\rho_{2}$  respectively. $\hat{s}_{13}=\pm{s}_{13}$ and a "+" corresponds to $\delta=0$ and a "-" sign to $\delta=\pi$. So, this equation will include all cases. Some cases work with the best fit values (BFV) of the oscillation parameters \cite{Gonzalez-Garcia:2014bfa}, while some cases work only with deviations from the BFV. We will allow some of these parameters to vary within a 3$\sigma$ range. Results are summarized in Table \ref{tab6}.
\begin{center}
  \begin{tabular}{|l|r|r|r|r|}
  \hline
  \multicolumn{1}{|c|}{\sffamily $\delta$}
  &\multicolumn{1}{|c|}{\sffamily p}&\multicolumn{1}{|c|}{\sffamily $\rho_{1}$}&\multicolumn{1}{|c|}{\sffamily $\rho_{2}$}&\multicolumn{1}{|c|}{\sffamily works with}\\
  \hline
  0&-&0&$\pi$&\makecell{IH, by best fit values of oscillation parameters,\\ $(m_{1},m_{2},m_{3})=(0.07613,0.07662,0.0585)$, $m_{\beta\beta}=0.0733$}\\
  \hline
  $\pi$&-&0&$\pi$&\makecell{IH, by best fit values of oscillation parameters,\\ $(m_{1},m_{2},m_{3})=(0.07635,0.07684,0.05878)$, $m_{\beta\beta}=0.07354$}\\
  \hline
  0&-&0&$\pi$&\makecell{NH, by best fit values of oscillation parameters,\\ $(m_{1},m_{2},m_{3})=(0.06353,0.06412,0.08058)$, $m_{\beta\beta}=0.06056$}\\
   \hline
  $\pi$&-&0&$\pi$&\makecell{NH, by best fit values of oscillation parameters,\\ $(m_{1},m_{2},m_{3})=(0.06315,0.06374,0.08028)$, $m_{\beta\beta}=0.0602$}\\
  \hline
  $\pi$&+&$\pi$&0&\makecell{IH, by best fit values of oscillation parameters,\\ $(m_{1},m_{2},m_{3})=(0.05735, 0.058, 0.03024)$, $m_{\beta\beta}= 0.02246$}\\
  \hline
  0&+&$\pi$&0&\makecell{IH, by best fit values of oscillation parameters,\\$(m_{1},m_{2},m_{3})=(0.04879,0.04955, 0.002516)$, $m_{\beta\beta}=0.0185$}\\
  \hline
  $\pi$&+&$\pi$&0&\makecell{NH, $\sin^2 \theta_{13}=0.0218$, $\sin^{2}\theta_{23}\in[0.382, 0.4]$, $m_{3}\in[0.12, 0.3]$, $\sin^2 \theta_{12}=[0.27, 0.297] $,\\ $m_{\beta\beta}\in[0.052, 0.14]$, $\sum m_{i}\in[0.34, 0.9]$}\\
  \hline
   0&+&$\pi$&$\pi$&\makecell{IH, $\sin^2 \theta_{13}=0.0218$, $\sin^{2}\theta_{23}\in[0.552, 0.644]$, $m_{3}\in[0, 0.002]$, $\sin^2 \theta_{12}=[0.313, 0.344] $,\\ $m_{\beta\beta}\in[0.0146, 0.0176]$}\\
  \hline
  \end{tabular}
  \captionof{table}{Results from $P_{6}$ type texture. "p" stands for a sign of a square root in (\ref{tpm1}). Masses are given in eVs.}
  \label{tab6}
  \end{center}
TYPE $P_{7}$
\\
For this case, the condition
$M_{\nu }^{(3,3)}=0$ gives:
\\
\beqs
\tan\theta_{12}=\frac{c_{23}s_{23}\hat{s}_{13}(m_{1}-m_{2}c_{1})}{m_{1}s_{23}^{2}+m_{2}c_{23}^{2}s_{13}^{2}c_{1}+m_{3}c_{23}^{2}c_{13}^{2}c_{2}}
\eeqs
\beq
\pm\frac{\sqrt{c^2_{23}s^2_{23}s^2_{13}(m_{1}-m_{2}c_{1})^2-(m_{1}s_{23}^{2}+m_{2}c_{23}^{2}s_{13}^{2}c_{1}+m_{3}c_{23}^{2}c_{13}^{2}c_{2})(m_{1}c_{23}^{2}s_{13}^{2}+m_{2}s_{23}^{2}c_{1}+m_{3}c_{23}^{2}c_{13}^{2}c_{2})}}{m_{1}s_{23}^{2}+m_{2}c_{23}^{2}s_{13}^{2}c_{1}+m_{3}c_{23}^{2}c_{13}^{2}c_{2}} \la{tpm2}
\eeq
Notations here are similar to those for case $P_6$ [see comment after Eq. (\ref{tpm1})]. Results are summarized in Table \ref{tab7}. As above, we have used data from  Ref.\cite{Gonzalez-Garcia:2014bfa}.
\\
\begin{center}
  \begin{tabular}{|l|r|r|r|r|}
  \hline
  \multicolumn{1}{|c|}{\sffamily $\delta$}
  &\multicolumn{1}{|c|}{\sffamily p}&\multicolumn{1}{|c|}{\sffamily $\rho_{1}$}&\multicolumn{1}{|c|}{\sffamily $\rho_{2}$}&\multicolumn{1}{|c|}{\sffamily works with}\\
  \hline
  0&+&$\pi$&0&\makecell{IH, by best fit values of oscillation parameters,\\ $(m_{1},m_{2},m_{3})=(0.9997, 0.10034,0.08729)$, $m_{\beta\beta}=0.04$}\\
  \hline
  0&-&0&$\pi$&\makecell{IH, $\sin^{2}\theta_{23}\in[0.389, 0.487]$, and bfv for remaining osc. parameters, \\$m_{3}\in[0.04496, 0.4138]$,  $m_{\beta\beta}\in[0.064, 0.398]$, $\sum m_{i}\in[0.178, 1.25]$}\\
   \hline
  $\pi$&+&$\pi$&0&\makecell{IH, by best fit values of oscillation parameters,\\  $(m_{1},m_{2},m_{3})=(0.05004, 0.05078,0.01142)$, $m_{\beta\beta}= 0.019$}\\
  \hline
  $\pi$&+&$\pi$&$\pi$&\makecell{IH, $\sin^{2}\theta_{23}\in[0.389, 0.448]$, $\sin^2 \theta_{12}=[0.325, 0.344]$ \\and bfv for remaining osc. parameters, $m_{3}\in[0, 0.001379]$, $m_{\beta\beta}\in[0.0146, 0.0165]$}\\
  \hline
  $\pi$&-&0&$\pi$&\makecell{IH, $\sin^{2}\theta_{23}\in[0.389, 0.488]$, and bfv for remaining osc. parameters, \\$m_{3}\in[0.04473, 0.6183]$, $m_{\beta\beta}\in[0.064, 0.59]$, $\sum m_{i}\in[0.178, 1.86]$}\\
  \hline
  0&+&$\pi$&0&\makecell{NH, $\sin^{2}\theta_{23}\in[0.621, 0.643]$, and bfv for remaining osc. parameters, \\$m_{3}\in[0.1246, 0.5928]$, $m_{\beta\beta}\in[0.046, 0.24]$, $\sum m_{i}\in[0.354, 1.77]$}\\
  \hline
  0&-&0&$\pi$&\makecell{NH, $\sin^{2}\theta_{23}\in[0.49, 0.643]$, and bfv for remaining oscillation parameters, \\$m_{3}\in[0.05803, 0.5187]$, $m_{\beta\beta}\in[0.0286, 0.4938]$, $\sum m_{i}\in[0.1196, 1.551]$}\\
  \hline
   $\pi$&-&0&$\pi$&\makecell{NH, $\sin^{2}\theta_{23}\in[0.49, 0.643]$, and bfv for remaining oscillation parameters, \\$m_{3}\in[0.05821, 0.5209]$, $m_{\beta\beta}\in[0.02895, 0.4959]$, $\sum m_{i}\in[0.1205, 1.558]$}\\
  \hline
  \end{tabular}
  \captionof{table}{Results from $P_{7}$ type texture. "p" stands for a sign of a square root in (\ref{tpm2}). Masses are given in eVs.}
  \label{tab7}
  \end{center}
\section{Relating cosmological CP and $\delta$}
\la{cosmology}
As we have already seen, from certain $2T_0Y_{32}$'s complex
phases cannot be factored out. Such couplings are:
$T_{4},T_{7},T_{9}$ and they give rise to complex mass matrices.
Here we calculate phase $\phi$ in terms of the CP phase entering in neutrino oscillation. Recall that the $\delta $  is predicted from the neutrino mass matrices (\ref{t124}),(\ref{t134}),(\ref{t237}),(\ref{t239}), which we have considered. Keeping in mind (\ref{pse}), we use  (\ref{nu1}) and (\ref{nu2}) to find the  numerical value of phase $\phi$ in each case.
\\
\\
Case of $M^{(12)}_{T_{4}}$ (Texture $P_1$):
\\
Equating (2,2), (3,3) and (2,3) matrix elements of both sides in Eq. (\ref{nu1}), we get the relations:
\beq
2a_{2}b_{2}|\bar m|e^{i\phi_{\bar m}}=e^{2i\omega_{2}}\mathcal{A}_{22},\quad 2a_{3}b_{3}e^{i\phi}|\bar m|e^{i\phi_{\bar m}}=e^{2i\omega_{3}}\mathcal{A}_{33},\quad (a_{3}b_{2}+a_{2}b_{3}e^{i\phi})|\bar m|e^{i\phi_{\bar m}}=e^{i(\omega_{2}+\omega_{3})}\mathcal{A}_{23}, \la{q0}
\eeq
with
\beq
\mathcal{A}_{ij}=U^{\ast}_{i1}U^{\ast}_{j1}m_{1}+U^{\ast}_{i2}U^{\ast}_{j2}m_{2}e^{i\rho_{1}}+U^{\ast}_{i3}U^{\ast}_{j3}m_{3}e^{i\rho_{2}}. \la{qq}
\eeq
Note, that from the neutrino sector all $\mathcal{A}_{ij}$ numbers are determined. Dividing the last relation in (\ref{q0}) in turn on the 1-st and 2-nd relations and then multiplying resulting two equations, we get the following relation:
\beq
xe^{i\phi}=\left(\frac{\mathcal{A}_{23}}{\sqrt{\mathcal{A}_{22}\mathcal{A}_{33}}}\pm \sqrt{\frac{\mathcal{A}^{2}_{23}}{\mathcal{A}_{22}\mathcal{A}_{33}}-1}\right)^{2}, \quad x\equiv \frac{a_{2}b_{3}}{a_{3}b_{2}}.
\eeq
Therefore, we have:
\beq
\phi ={\rm Arg}\left[\left(\frac{\mathcal{A}_{23}}{\sqrt{\mathcal{A}_{22}\mathcal{A}_{33}}}\pm \sqrt{\frac{\mathcal{A}^{2}_{23}}{\mathcal{A}_{22}\mathcal{A}_{33}}-1}\right)^{2}\right].
\eeq
From here, using results given in Table \ref{tab1}, we find numerical value of $\phi$:
\beq
\phi=\pm1.287.
\eeq
\\
In a pretty similar way, for remaining three neutrino mass matrices (\ref{t134}),(\ref{t237}),(\ref{t239}), for the phase $\phi$ we get:
\\
\beq
\phi ={\rm Arg}\left[\left(\frac{\mathcal{A}_{23}}{\sqrt{\mathcal{A}_{22}\mathcal{A}_{33}}}\pm \sqrt{\frac{\mathcal{A}^{2}_{23}}{\mathcal{A}_{22}\mathcal{A}_{33}}-1}\right)^{2}\right], \quad \phi ={\rm Arg}\left[\left(\frac{\mathcal{A}_{13}}{\sqrt{\mathcal{A}_{11}\mathcal{A}_{33}}}\pm \sqrt{\frac{\mathcal{A}^{2}_{13}}{\mathcal{A}_{11}\mathcal{A}_{33}}-1}\right)^{2}\right],
\eeq
\beq
\phi={\rm Arg}\left[\left(\frac{\mathcal{A}_{12}}{\sqrt{\mathcal{A}_{11}\mathcal{A}_{22}}}\pm \sqrt{\frac{\mathcal{A}^{2}_{12}}{\mathcal{A}_{11}\mathcal{A}_{22}}-1}\right)^{2}\right],
\eeq
which yield
\beqs \phi=\pm1.169,\quad \phi^{\rm NH}=\pm2.957\quad {\rm and}\quad\phi^{\rm IH}=\pm3.124,\eeqs \beq
\phi^{\rm NH}=\pm3.058\quad{\rm and}\quad \phi^{\rm IH}=\pm3.136
\eeq
respectively. For these we have used results
given in Tables: \ref{tab2}, \ref{tab3'} and \ref{tab4} resp.
Note, that $\phi$ phases in all four cases  have been found  for
the reason that with a predictive neutrino sector there is no
undetermined parameter. This makes the whole scenario
really attractive to study the baryon asymmetry via the
leptogenesis (for similar studies see: \cite{Frampton:2002qc},\cite{Frampton:2002yf},\cite{{Babu:2007zm},{Babu:2008kp},{Harigaya:2012bw},{Ge:2010js}},\cite{Frampton:2004df}). As mentioned, since the $\phi$ participates in the coupling of RHN states with $l$ and $h_{u}$ (\ref{r21}) it will control CP asymmetric decays of the N states. Thus, it is interesting to look into the details of the leptogenesis within the scenarios we have considered here. This will be pursued in a subsequent work \cite{AZ}.
\section{Conclusions}
\la{conclusion}
Within the MSSM augmented with two quasi-degenerate right-handed neutrinos, we analyzed all possible two texture zero $3\times 2$ Yukawa matrices, which together with minimal $\rm d=5$ operator couplings contribute to the light neutrino mass matrices. All viable neutrino mass matrices have been investigated and predictive relations were derived. Cosmological CP violation has been related to the leptonic CP violating $\delta $ phase.
Further work will be focused on details of realizations of resonant leptogenesis. It is also desirable to get texture zeros with the help of flavor symmetries in a spirit of Refs.\cite{Fritzsch:2011qv},\cite{Babu:2007zm},\cite{{Pakvasa:1977in},{Binetruy:1996xk},{Lola:1998xp},{Vissani:1998xg},{Shafi:1998dv},{Barbieri:1999km},{Berezhiani:2000cg},{Ma:2001dn},{Chkareuli:2001dq},{Babu:2004tn},{Hagedorn:2006ug},{Nandi:2007cw},{King:2013eh}}.
These and related issues will be addressed elsewhere.
\subsubsection*{Acknowledgments}
Research of Z.T. is partially supported by Shota Rustaveli National Science Foundation (Contracts No. 31/89 and No. DI/12/6-200/13).
\bibliographystyle{unsrt}

\end{document}